\documentclass[useAMS,usenatbib,usegraphicx]{mn2e}
\usepackage{subfigure}

\title[The effect of Lyman $\alpha$ radiation on mini-Neptune atmospheres around M stars]{The effect of Lyman $\alpha$ radiation on mini-Neptune atmospheres around M stars: application to GJ 436b}

\author[Yamila Miguel, Lisa Kaltenegger, Jeffrey L. Linsky and Sarah Rugheimer]{Yamila Miguel$^{1}$\thanks{miguel@mpia.de}, Lisa Kaltenegger$^{1,2}$, Jeffrey L. Linsky$^{3}$ and Sarah Rugheimer$^{2}$\\
$^{1}$Max Planck Institut f\"ur Astronomie, K\"onigstuhl 17, 69117, Heidelberg, Germany\\
$^{2}$Harvard Smithsonian Center for Astrophysics, 60 Garden St.,
  Cambridge, MA, 02138 USA\\
$^{3}$JILA, University of Colorado and NIST, 440 UCB Boulder, CO 80309-0440, USA}

\begin{document}

\pagerange{\pageref{firstpage}--\pageref{lastpage}} 

\label{firstpage}

\maketitle

\begin{abstract}
Mini-Neptunes orbiting M stars are a growing population of known
exoplanets. Some of them are located very close to their host star,
receiving large amounts of UV radiation. Many M stars emit strong
chromospheric emission in the H I Lyman $\alpha$ line (Ly$\alpha$) at
1215.67~\AA, the brightest far-UV emission line. We show that
the effect of incoming Ly$\alpha$ flux can significantly change the
photochemistry of mini-Neptunes' atmospheres. We use GJ~436b as an
example, considering different metallicities for its atmospheric
composition. For solar composition, H$_2$O-mixing ratios show the
largest change because of Ly$\alpha$ radiation. H$_2$O absorbs most of
this radiation, thereby shielding CH$_4$, whose dissociation is driven
mainly by radiation at other far-UV wavelengths ($\sim1300$~\AA). 
H$_2$O photolysis also affects other species in the atmosphere, 
including  H, H$_2$, CO$_2$, CO, OH and O. For an atmosphere with 
high metallicity, H$_2$O- and CO$_2$-mixing ratios show the biggest 
change, thereby shielding CH$_4$. Direct measurements of the UV flux 
of the host stars are important for understanding the photochemistry 
in exoplanets' atmospheres. This is crucial, especially in the 
region between 1 and 10$^{-6}$ bars, which is the part of the 
atmosphere that generates most of the observable spectral features.
\end{abstract}


\section{INTRODUCTION}

Recent exoplanet surveys have discovered the first planets with sizes
between 2 Earth radii (R$_{\rm Earth}$) and 3.5$R_{\rm Earth}$, 
the size of Neptune. These planets, known as mini-Neptunes, revolve around
the M stars GJ 436b \citep{bu04}, GJ 1214b
\citep{cha09}, Kepler 26b and Kepler 32d \citep{bo11} and GJ 3470b
\citep{bo12}. Since mini-Neptunes around M stars are expected to be
abundant \citep{la04} and M stars are the most common stars in the
solar neighborhood \citep{ch03}, we expect that many more
mini-Neptunes will be discovered in the near future.   

Since M stars have low effective temperatures, the bulk of
their flux is emitted in the optical and near IR. Inactive M stars have
very low-photospheric--UV-continuum emission when compared to
solar-type stars. Direct observations of active M dwarfs show a high
flux in the far-UV (FUV 912--1700~\AA) \citep{ev14}, and the percentage of total UV
flux from the star in the H I Lyman $\alpha$ line (Ly$\alpha$) is between
37 and 75 per~cent compared to 0.04 per~cent for the Sun \citep{fr13}. 
While most of the stellar Ly$\alpha$ radiation is scattered or
absorbed in the interstellar medium, Ly$\alpha$ is the brightest
FUV emission line in the stellar spectrum seen by an exoplanet
\citep{fr13} which makes this emission line critical for atmospheric 
photochemistry.

While the effect of extreme-UV irradiation (EUV 200--911~\AA) was
studied for Earth-like \citep{la11} and giant planets
\citep{ye04,la03,le04,ye06,mu09,ko10,sa11,le12} and the consequence
of high FUV M star irradiation was explored for habitable planets'
atmospheres \citep{sc07,bu07,anti10}, the effect of FUV irradiation
and especially Ly$\alpha$ flux on hot mini-Neptune atmospheres has not
yet been evaluated. Solar Ly$\alpha$ radiation is known to have
a strong impact on the
photochemistry of the planets in our Solar System, and the effects of
stellar Ly$\alpha$ radiation on the photochemistry of hot extrasolar planets 
are expected to be important, but such effects have not yet been quantified.
In particular, the effect of Lyman-$\alpha$ radiation on
the thermal profile in mini-Neptune atmospheres is only beginning to
be studied \citep{lav11}. 

We have performed 1D simulations of hot mini-Neptune atmospheres under 
different irradiation conditions to make a deeper exploration of the
effects of Ly$\alpha$ flux on the photochemistry. We focus
here on the mini-Neptune GJ 436b as an example, using recent UV 
observations including the reconstructed Ly$\alpha$ flux \citep{fr13}. 

\section{MODEL DESCRIPTION}\label{model}
\subsection{Stellar flux}\label{stellar-flux}

GJ 436 is an M3 dwarf (T$_{{\rm eff},\star}=3416$~K) with a radius 
$R_{\star}=0.455 R_{\odot}$ \citep{kaspar12}. Its coronal flux
(log$L_X=27.16\pm0.34$) is smaller than the mean for M dwarfs
(log$L_X=27.6$), indicating a low activity corona for GJ 436
\citep{po10}. The first reconstructed Ly$\alpha$ emission line
  profile performed for GJ 436 was based on (HST/STIS) observations
  \citep{eh11}. Here we use the most recent UV spectral observation of
  GJ 436 from 1150 to 3140~\AA$~$\citep{fr13}, including the
reconstructed stellar Ly$\alpha$ emission line profile, which is 1750
ergs cm$^{-2}$ s$^{-1}$ at 0.03AU \citep{fr13, lins13}. The spectrum
is available in the MUSCLES (Measurements of the Ultraviolet Spectral
Characteristics of Low-mass Exoplanetary Systems)
website\footnote{http://cos.colorado.edu/∼kevinf/muscles.html}. We
merged those UV observations with optical synthetic spectra between
2800 and 45450~\AA\ from the PHOENIX models \citep{all07}. We adopt 
[Fe/H]=0.04 \citep{ro12}, $V sin(i)<1$~km~s$^{-1}$ \citep{ma92} and 
Log(g)=4.83 \citep{ma07}.

\begin{figure*}
 \begin{center}
\subfigure[][]{\label{flux}\includegraphics[angle=90,width=.7\textwidth]{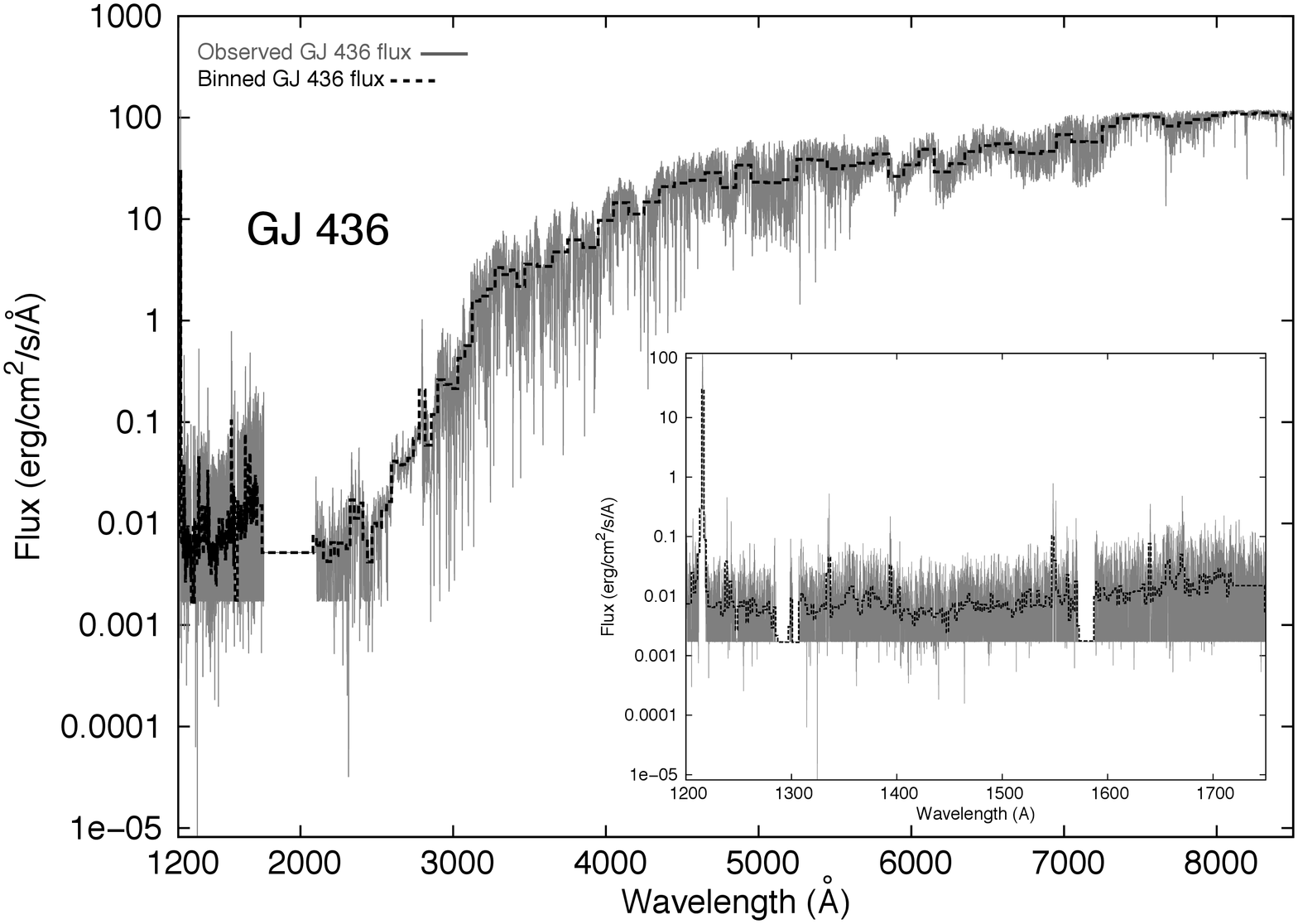}}
\subfigure[][]{\label{high}\includegraphics[angle=0,width=.34\textwidth]{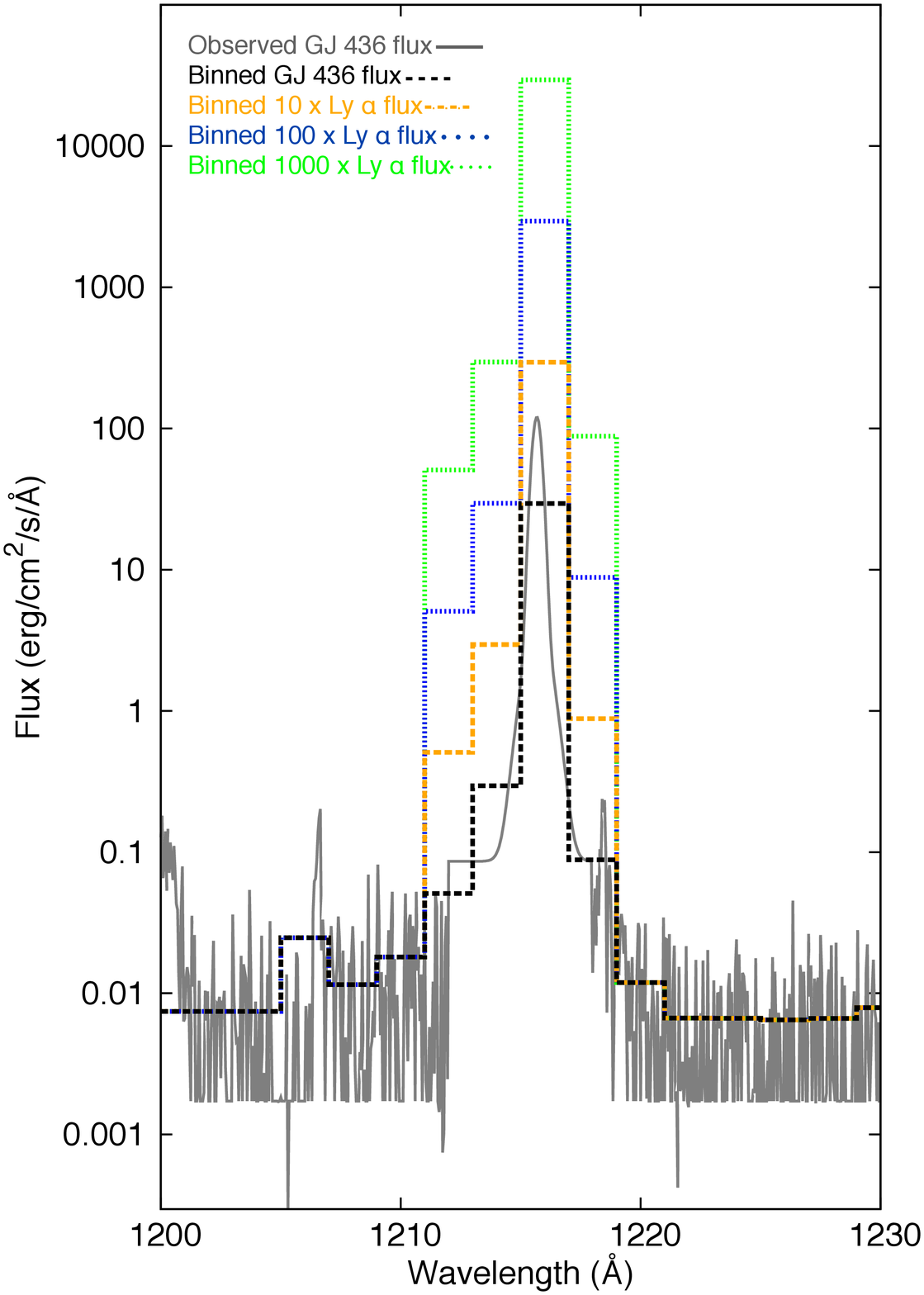}}\subfigure[][]{\label{low}\includegraphics[angle=0,width=.34\textwidth]{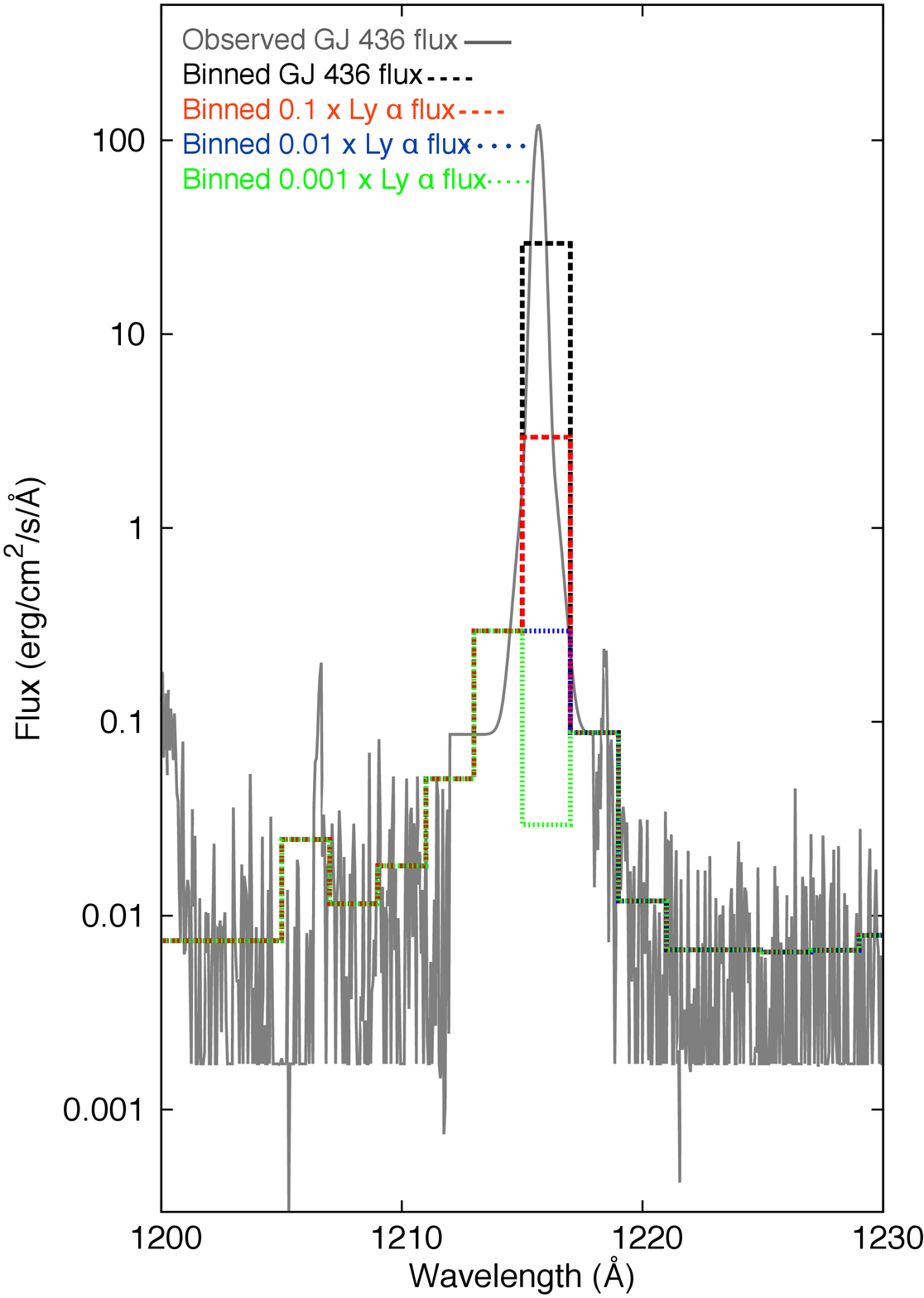}}
\end{center}
\caption{Flux of GJ 436 vs. wavelength obtained from UV direct
  measurements + model (grey), with the binning used in the code
  (black line). Flux between 1200 and 1800~\AA\ is highlighted (right
  bottom panel). Figs. \ref{high} and \ref{low} show the different
  Ly$\alpha$ flux scenarios explored in the paper. Fig. \ref{high}:
  1000 x Ly$\alpha$  (green dotted), 100 x Ly$\alpha$  (blue dotted)
  and 10 x Ly$\alpha$ (orange dots and dashes line).  Fig. \ref{low}:
  0.1 x Ly$\alpha$ (red dashed), 0.01 x Ly$\alpha$ (blue dotted) and
  0.001 x Ly$\alpha$ (green dotted line). The flux values are
  normalized to 1AU.} 
\label{fluxes}
\end{figure*}

The resulting GJ 436 flux for $\lambda<$ 8550~\AA$~$ is shown in
Fig. \ref{flux} (grey solid), with the black line showing the
binned-stellar input used in our code at a resolution of 2~\AA$~$in
the FUV and 16 to 50~\AA $~$ from 2100 to 
45450~\AA. Note that we use the stellar flux between 1200 and
8550~\AA$~$in our photochemical calculations. To analyze the importance of
the Ly$\alpha$ flux in exoplanet atmospheres, we model 4 scenarios
shown in Fig. \ref{high}: GJ 436b as irradiated by a star with 1000 x
Ly$\alpha$
(green), 100 x Ly$\alpha$ (blue), 10 x Ly$\alpha$ (orange), and
1 x Ly$\alpha$ (black), where n x Ly$\alpha$ is the
reconstructed Ly$\alpha$ flux for GJ 436 multiplied by the factor n. 
Since Ly$\alpha$ radiation could be absorbed by an exoplanet's upper 
atmosphere \citep{lav11},
we also explore three scenarios where Ly$\alpha$ flux is reduced to 
0.1 x Ly$\alpha$ (red), 0.01 x Ly$\alpha$ (blue) and 0.001 x Ly$\alpha$ 
(green line), as shown in
Fig. \ref{low}. Note that all M dwarf stars have chromospheres with a
wide range of activity, and our calculations explore the effects of
extreme differences in Ly$\alpha$ fluxes on exoplanet
atmospheres. One thousand times the Ly$\alpha$ flux of GJ 436 corresponds to
 a very active M dwarf star, and 0.001 x Ly$\alpha$ corresponds to
 very strong absorption of that radiation in the planet's extended upper
 atmosphere. We have not considered variations in the total FUV flux,
 which will scale with Ly$\alpha$ flux, in order to isolate the
 effects of the Ly$\alpha$ line which dominates the total UV flux. 

\subsection{Atmospheric modeling}\label{atmospheres}

GJ 436b is a mini-Neptune with $0.078~M_{\rm Jupiter}$ and 
$0.369~R_{\rm Jupiter}$, orbiting its host star at 0.03~AU
\citep{kaspar12}. We use a 1D model to calculate GJ 436b's atmospheric
thermal structure and disequilibrium chemistry including molecular
diffusion, vertical mixing and photochemistry \citep{mi14}.

We updated and improved the resolution of the cross sections for the
different molecules in the FUV compared to previous studies
\citep{ravi,mi14}. 
Our new cross-section database includes high spectral-resolution measurements of absorption cross sections of high temperature CO$_2$ \citep{ve13}, O$_2$,  CH$_4$ and H$_2$O from the MPI-Mainz--UV-VIS Spectral Atlas of Gaseous Molecules \footnote{www.uv-vis-spectral-atlas-mainz.org}, giving preference to the higher spectral-resolution measurements data (see also \citet{yan13}). Note that due to the lack of data at longer wavelengths, the H$_2$O cross section was extrapolated between $1900$ and $2400$~\AA$~$\citep{ravi} (black dotted lines in Fig. \ref{cross-section}).

The 1D atmospheric chemistry model (see \citet{mi14} and \citet{ravi}
for a detailed explanation) considers equilibrium chemistry for higher
temperatures and pressures and chemical disequilibrium in the upper
atmosphere where the densities are low, and these processes have
shorter timescales. The code includes disequilibrium processes
(molecular diffusion, vertical mixing and photochemistry) and focuses
on the following species in 179 reactions: O, O(1D), O$_2$, H$_2$O, H,
OH, CO$_2$, CO, HCO, H$_2$CO, CH$_4$, CH$_3$, CH$_3$O, CH$_3$OH, CH,
CH$_2$, H$_2$COH, C and H$_2$ \citep{ravi,mi14}. Other important
species like HCN and C$_2$H$_2$ are beyond the scope of this work and
will be included in a future study. To characterize
  vertical-mixing processes in the atmosphere, we use a constant-eddy
  diffusion coefficient. Although this coefficient is difficult to
  constrain, results derived from comparisons between 1D and 3D models
  for HD 209458b show that the eddy-diffusion coefficient has values
  between $K_{ZZ}=10^8$ and $K_{ZZ}=10^{12}$~cm$^2$~s$^{-1}$
  \citep{pa13}. Following these results, we adopt an intermediate
  value of 
$K_{ZZ}=10^9$~cm$^2$~s$^{-1}$ in our calculations. Note that this value 
is not known for most hot extrasolar planets, and different values
lead to differences in the upper atmosphere-mixing ratios
\citep{vi11,mi14}. An exploration of extreme cases was performed by
\citet{mi14} who showed that dissociation becomes less efficient as 
vertical mixing becomes stronger in the atmosphere which affects the 
atmospheric abundances and the effect of Ly$\alpha$ radiation in the 
atmosphere.

We obtained the temperature and pressure profiles from a radiative
atmosphere model developed by for highly irradiated exoplanets by 
\citet{gu10} who found global mean thermal profiles comparable to 
detailed-atmospheric model calculations. The temperature structure of the exoplanet atmosphere as a function of the mean optical depth for thermal radiation ($\tau$) is given by equation (\ref{T}) \citep{gu10}:

\begin{displaymath}
T_p^4=\frac{3T_{int}^4}{4}\bigg(\frac{2}{3}+\tau\bigg)+
\end{displaymath}
\begin{equation}\label{T}
\frac{3T_{irr}^4}{4} f \bigg(\frac{2}{3}+\frac{1}{\gamma \sqrt{3}}+\bigg(\frac{\gamma}{\sqrt{3}}-\frac{1}{\gamma \sqrt{3}}\bigg)e^{-\gamma \tau \sqrt{3}}\bigg).
\end{equation}
We use $\gamma= 0.05$ which provides a good match to the
  thermal structures retrieved from observations of GJ 436b
  \citep{st10,madu11,mo13} and GCMs models \citep{le10}. The planet's
internal temperature (T$_{\rm int}$) is usually adopted as 50~K for
  old mini-Neptunes [e.g. \citet{mill11}], but GJ 436b has a high
  eccentricity 
[e=0.146 \citet{kaspar12}] which implies a potentially tidally heated
  exoplanet \citep{ag14}. We therefore adopt T$_{\rm int}=300$~K [following \citet{gu10}]. Different values of T$_{int}$ might change the thermal profile deep in the atmosphere \citep{mo13,ag14}.

\begin{figure}
\includegraphics[angle=90,width=.48\textwidth]{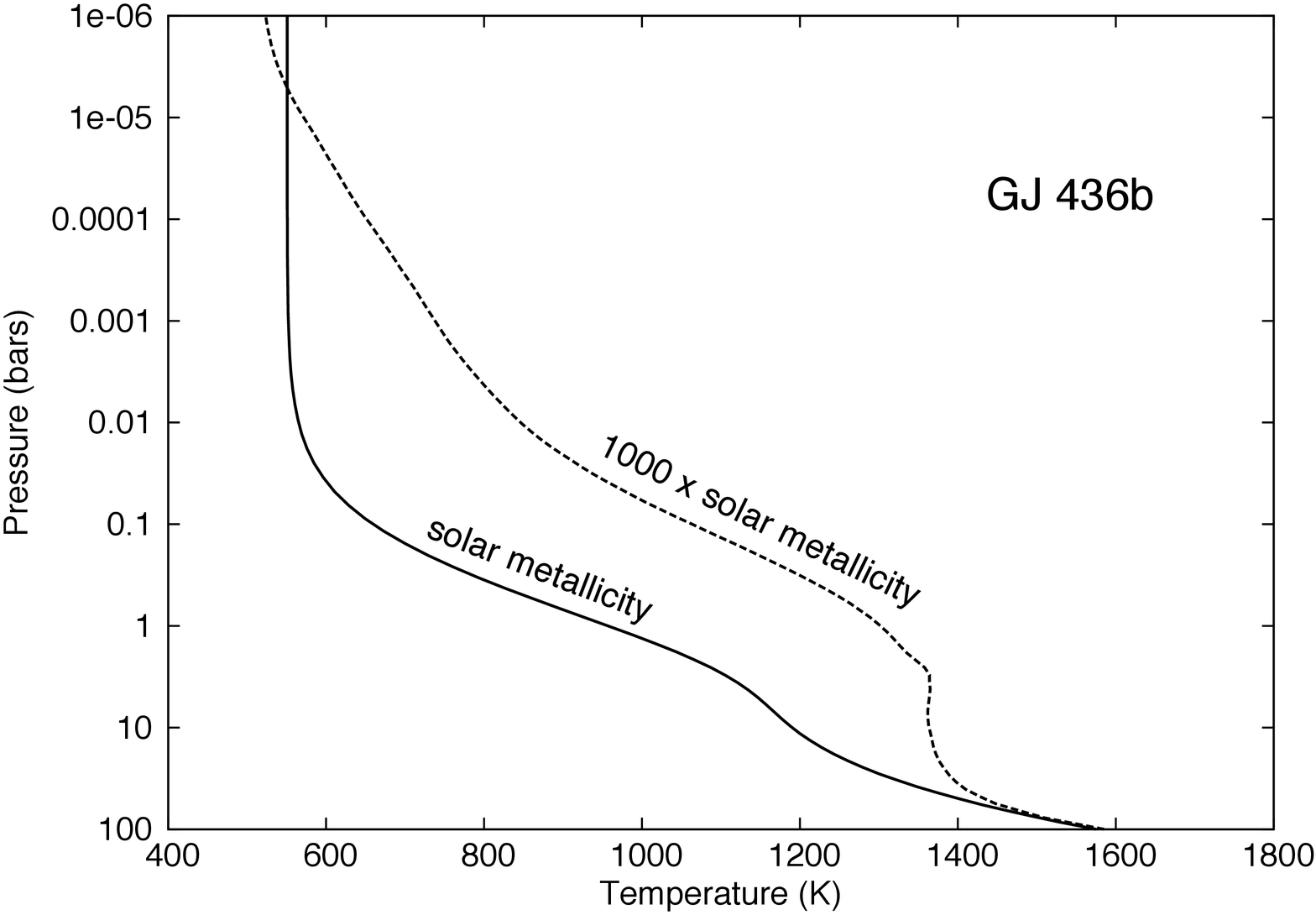}
\caption{Temperature-pressure profile for GJ 436b for solar (solid) and 1000 x solar metallicity (dashed line) in the atmosphere.}
\label{TP}
\end{figure}

We use 100 atmospheric layers from $5\times10^{-8}$ to 200~bars. The
thermal profile for GJ 436b is shown in Fig. \ref{TP} for solar
metallicity (solid) and 1000 x solar enrichment \citep{mo13}  (dashed
line). The 1000 x solar metallicity thermal profile was
  derived in \citet{mo13} using a PHOENIX model with the assumption that the stellar heating causes a circulation that inefficiently redistributes energy to the night side, as described in \citet{bar01,bar05}. 

\section{RESULTS: EFFECT OF Ly$\alpha$ FLUX ON MINI-NEPTUNE ATMOSPHERES}\label{results}

Photodissociation of atmospheric molecules occurs mostly in the upper
atmosphere. This process is typically a one-way reaction which means
that the 
probability of recombination and subsequent photon emission is
negligible, thereby maintaining chemical disequilibrium in this region 
of the atmosphere.

All major species in mini-Neptune atmospheres considered in our model
have a dissociation energy corresponding to wavelengths
shorter 
than 2800~\AA$~$--- H$_2$O (dissociation energy 5.17~eV, equivalent to
2398~\AA), CO$_2$ (5.52~eV or 2247~\AA), CH$_4$ (4.55~eV or 2722~\AA),
H$_2$ (4.52~eV or 
2743~\AA) and CO (11.14~eV or 1113~\AA). Therefore, UV radiation dominates the photochemistry in mini-Neptune atmospheres. 

\begin{figure}
\subfigure[][]{\label{cross-sectionA}\includegraphics[angle=90,width=.48\textwidth]{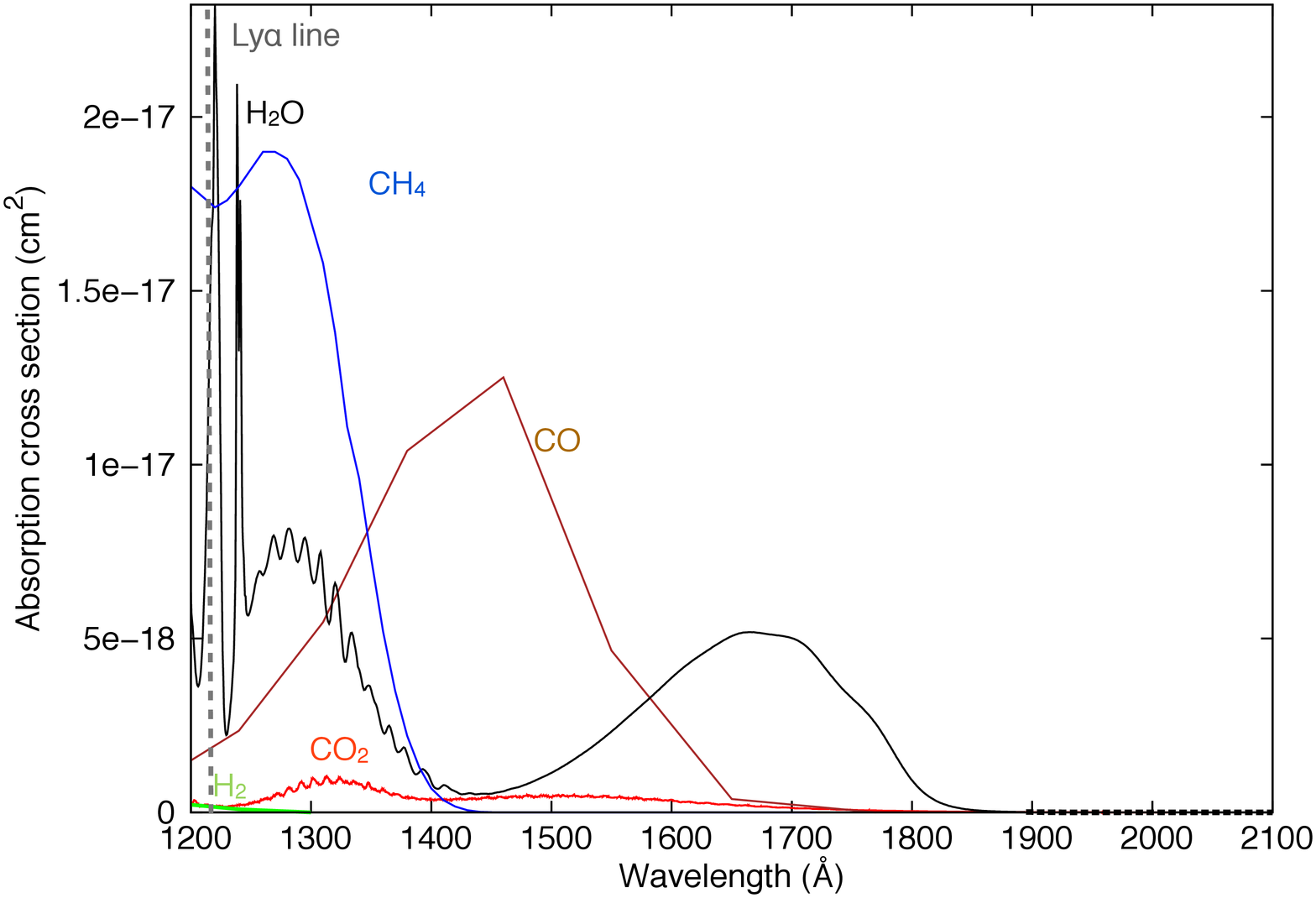}}
\subfigure[][]{\label{cross-sectionB}\includegraphics[angle=90,width=.48\textwidth]{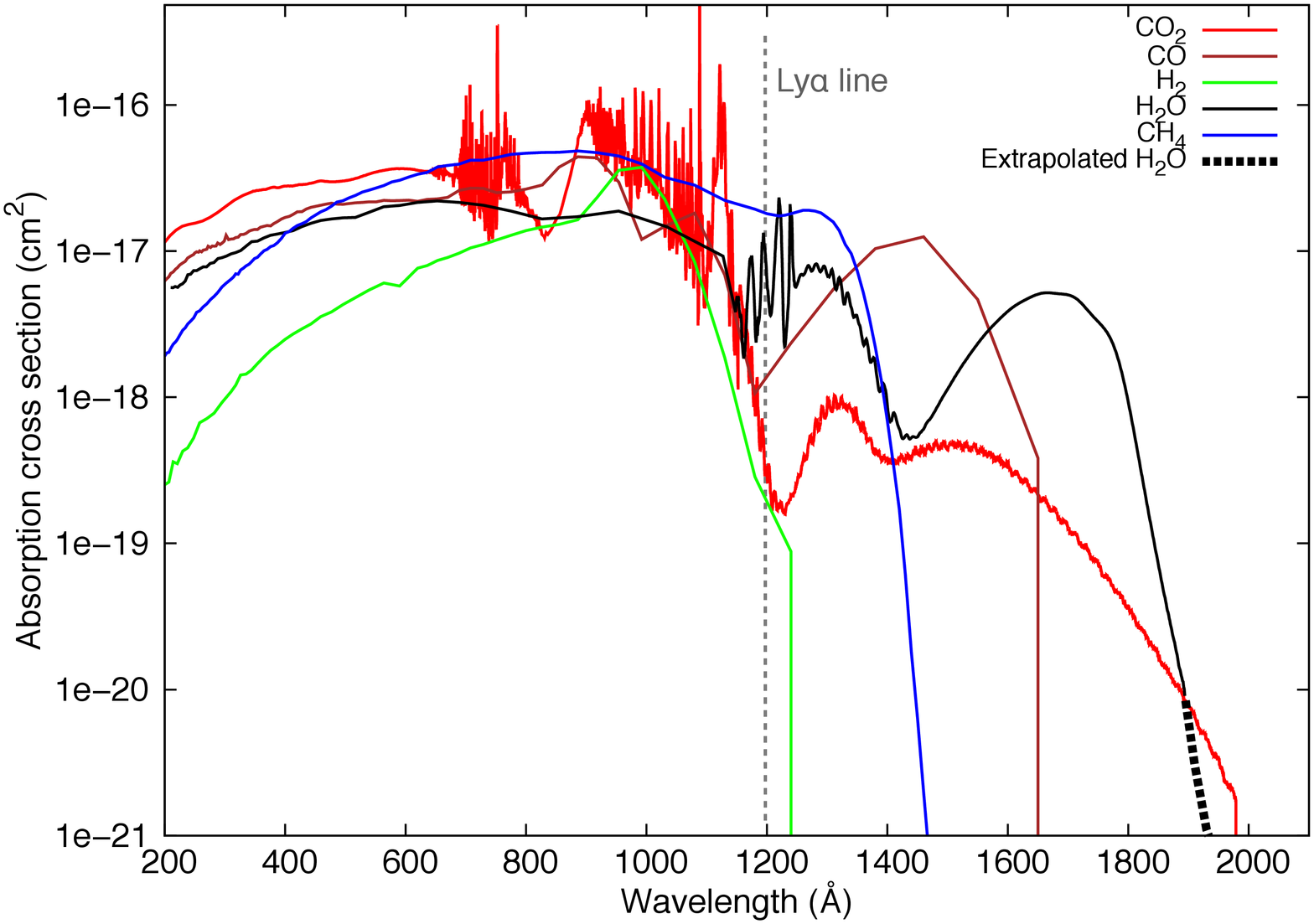}}
\caption{FUV cross sections vs. wavelength of some important molecules
  in GJ 436b's atmosphere: H$_2$ (green), H$_2$O (black), CH$_4$
  (blue), CO (brown) and CO$_2$ (red). Extrapolation of H$_2$O cross
  section beyond 1900~\AA$~$is shown (black dotted line). The Ly$\alpha$
  line is shown with grey-dashed line. The region 
between 1200 and 2100~\AA$~$is shown in Fig. \ref{cross-sectionA}, and 
between 200 and 2100~\AA$~$is shown in Fig. \ref{cross-sectionB}. }
\label{cross-section}
\end{figure}

Fig. \ref{cross-section} shows the absorption cross sections as a
function of wavelength for the most abundant molecules in the mini-Neptune atmospheres considered in this paper: H$_2$ (green) \citep{ba76}, H$_2$O (black) \citep{mo05}, CO (brown) \citep{su55}, CH$_4$ (blue) \citep{lee01,ch04} and CO$_2$ (red) \citep{hu10,ve13}. Fig. \ref{cross-sectionA} shows the region between 1200 and 2100~\AA, and a larger region (between 200 and 2100~\AA) is shown in Fig. \ref{cross-sectionB}. 

Fig. \ref{cross-section} shows that H$_2$O and CH$_4$ have higher
cross sections between 1000 and 1400~\AA\ and are especially
susceptible to Ly$\alpha$ radiation from the host M star. Because of
its short wavelength for dissociation, CO does not dissociate in
response to FUV radiation, but has two peaks at 1332 and 1474~\AA\
corresponding to the excitation of the various excited states of the
molecule. As seen in the right top panel of Fig. \ref{cross-section},
CO$_2$ is a strong absorber at short wavelengths ($\lambda<1200$~\AA,
maximum at 900~\AA), 
but CO$_2$ does not have a high cross section in the FUV. 
Even though H$_2$ has a small cross section in the FUV, it becomes a
strong absorber at short wavelength ($\lambda<1200~$\AA) in exoplanet
atmospheres because of its high abundance. Thus H$_2$ 
shields other molecules from very shortwave radiation. 

EUV radiation
is not included here because it is absorbed in the upper atmosphere
and does not reach pressures in the range described in this
paper. Note that the fate of Ly$\alpha$ photons as they travel through
the extended H-rich thermosphere is 
not yet clear as this radiation could be significantly scattered by H
in the planet's extended thermosphere \citep{lav11,ko10} (see Section
\ref{discussion}). A more detailed model of the extended upper
atmosphere and atmospheric escape is beyond the scope of this paper.

Observations of GJ 436b's atmosphere indicate a CO-rich and
CH$_4$-poor atmosphere \citep{st10,madu11,kn14}. This composition 
can be explained
by adopting high metallicities, between 230 to 2000 x solar
\citep{mo13}. We therefore explore the effect of high FUV, and especially Ly$\alpha$, flux on mini-Neptune atmospheres, for solar (Section \ref{solar}) as well as high metallicity (Section \ref{1000xsolar}) atmospheric composition.   

\subsection{Effect of high Ly$\alpha$ flux}
\subsubsection{Atmospheres with solar metallicities}\label{solar}

\begin{figure*}
  \begin{center}
\subfigure[][]{\label{photolysis}\includegraphics[angle=90,width=.45\textwidth]{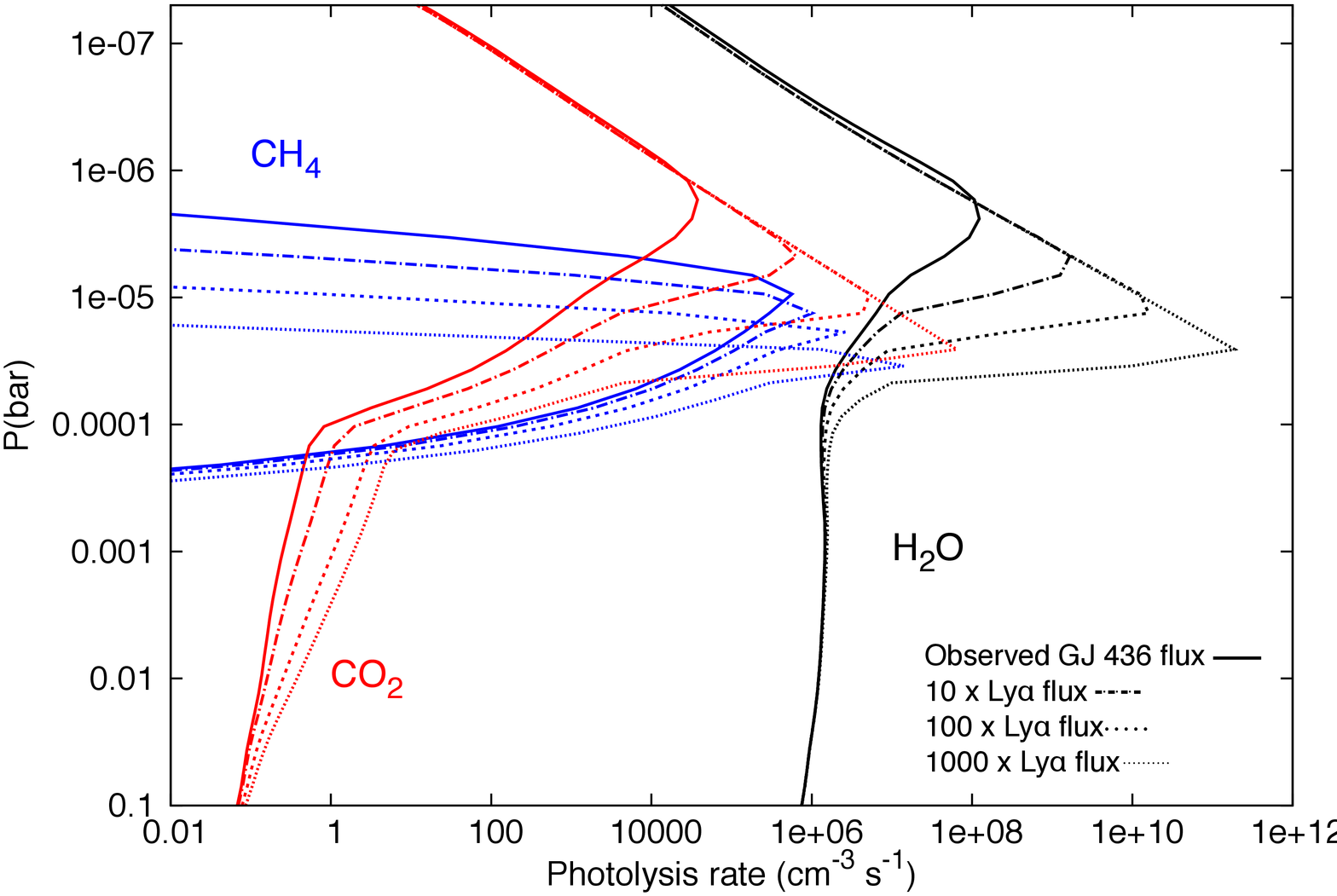}}
\subfigure[][]{\label{photolysis-frequency}\includegraphics[angle=90,width=.45\textwidth]{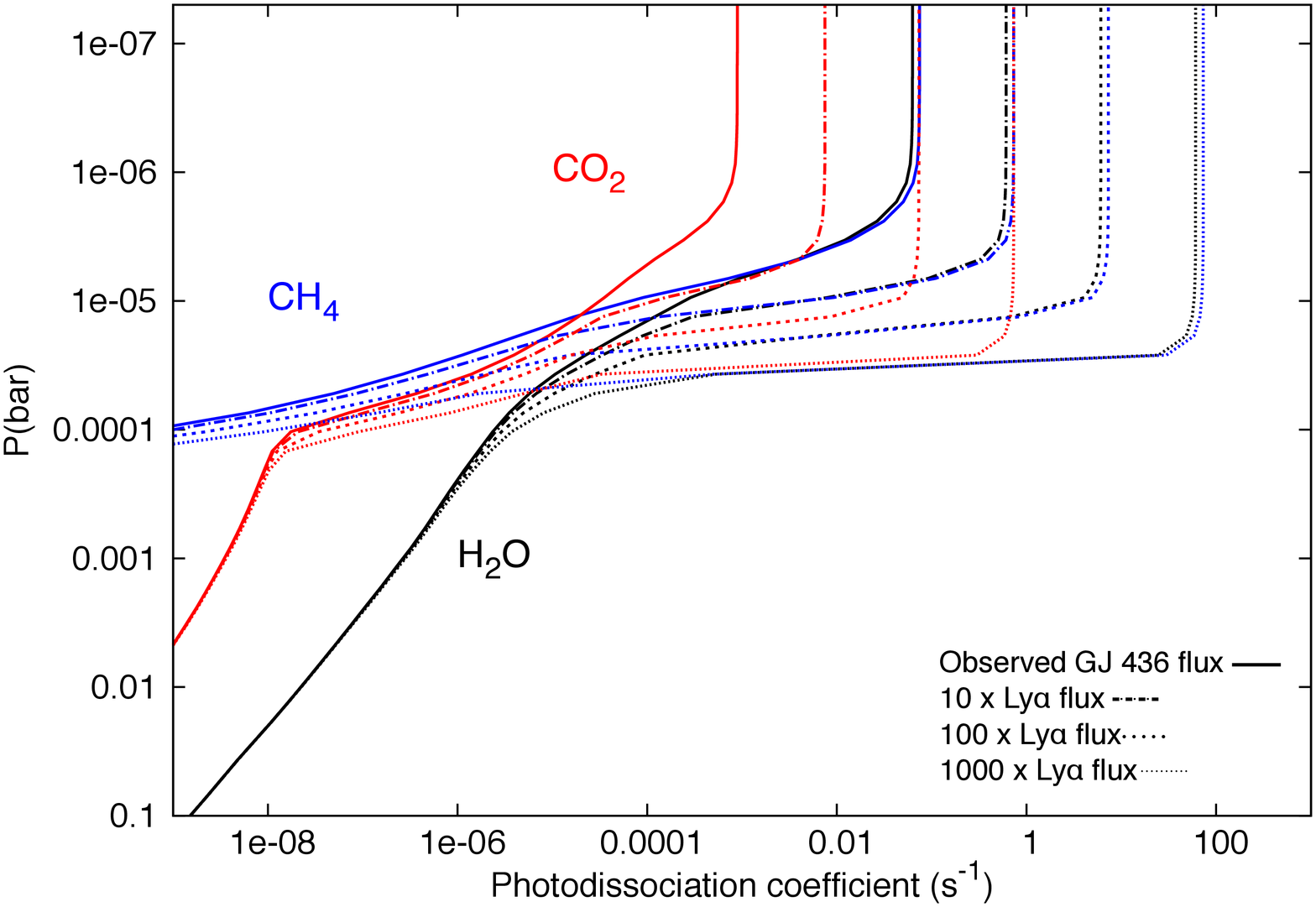}}
\subfigure[][]{\label{photolysis-metal}\includegraphics[angle=90,width=.45\textwidth]{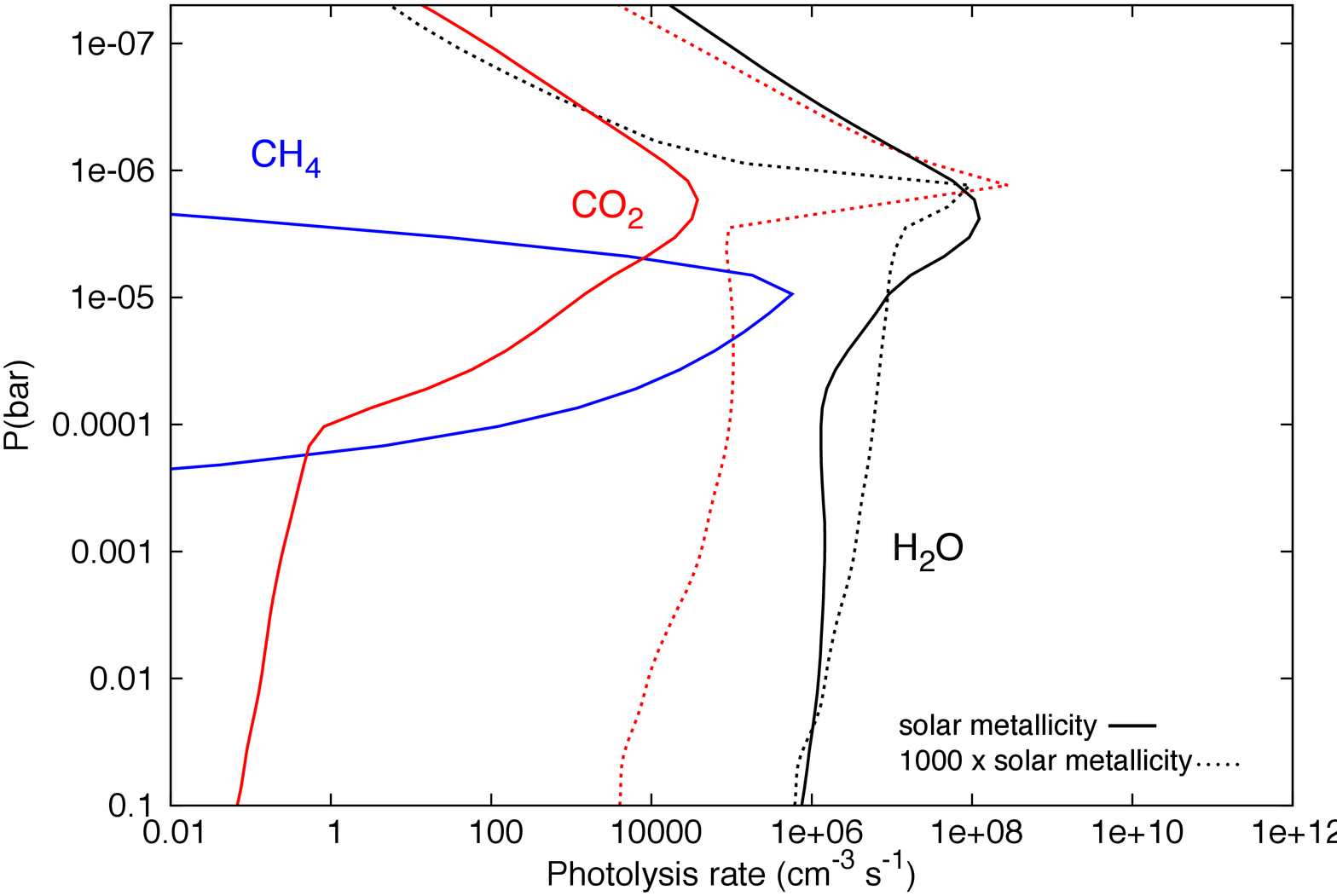}}
\subfigure[][]{\label{photolysis-metal-frequency}\includegraphics[angle=90,width=.45\textwidth]{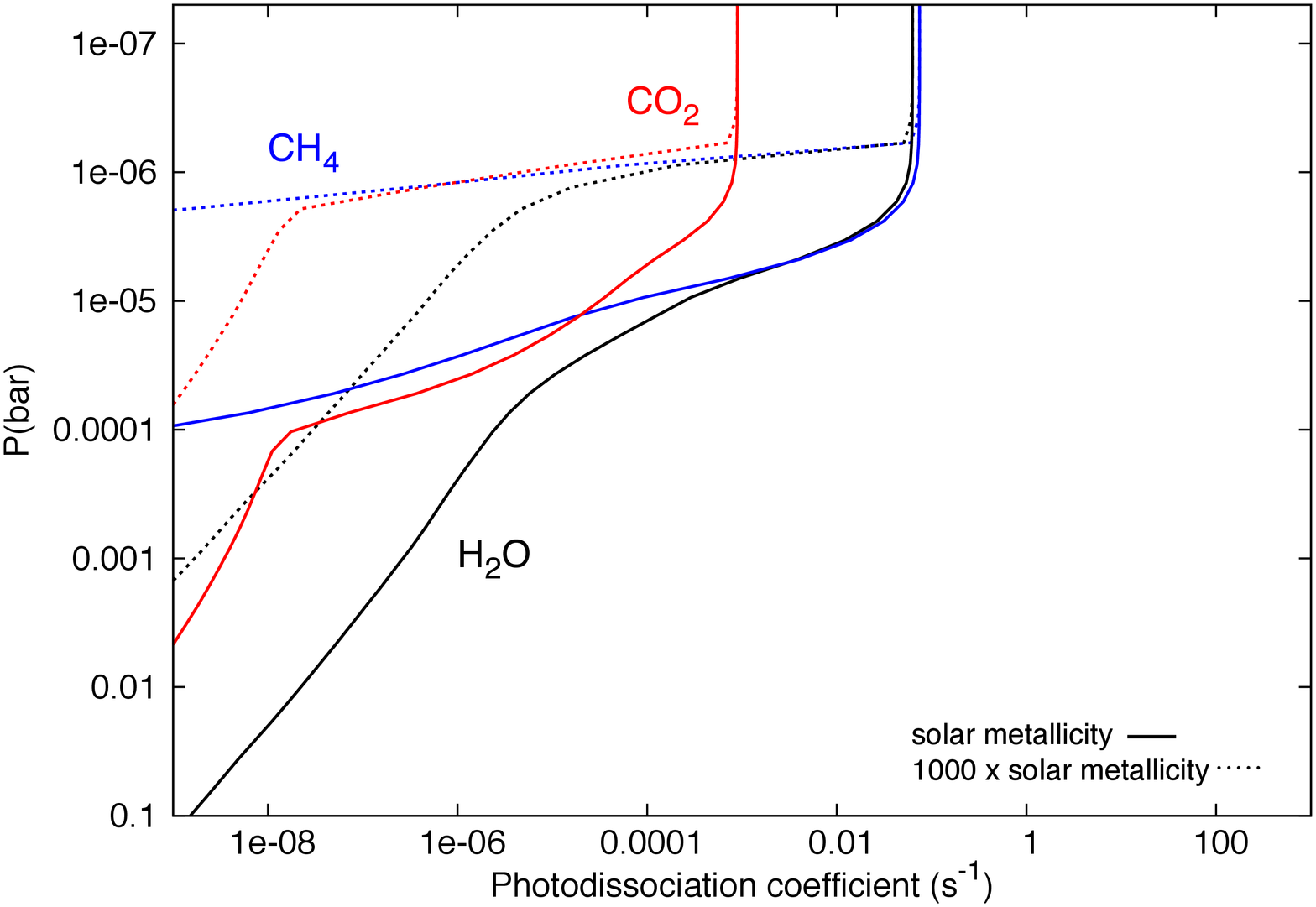}}
 \end{center}
  \caption{Photolysis rates (left panels) and photodissociation
  coefficient (right panels) vs. pressure in the atmosphere of GJ~436b
  for H$_2$O (black), CO$_2$ (red) and CH$_4$ (blue).  Top panels:
  \textit{solar composition} for four Ly$\alpha$ flux scenarios: 1000
  x Ly$\alpha$ (small dots), 100 x Ly$\alpha$ (dotted), 10 x
  Ly$\alpha$ (dots and dashes) and the reconstructed Ly$\alpha$ flux (solid line). Bottom panels: comparison between solar composition (solid) and 1000 x solar metallicity atmosphere (dotted lines) for the \textit{reconstructed Ly$\alpha$ flux}. Note that solid lines in Figs. \ref{photolysis-metal} -- \ref{photolysis-metal-frequency}  are the same as in Fig. \ref{photolysis} -- \ref{photolysis-frequency}, respectively.}
\end{figure*}

\begin{figure*}
  \begin{center}
\subfigure[]{\label{lyman-1}\includegraphics[angle=90,width=.45\textwidth]{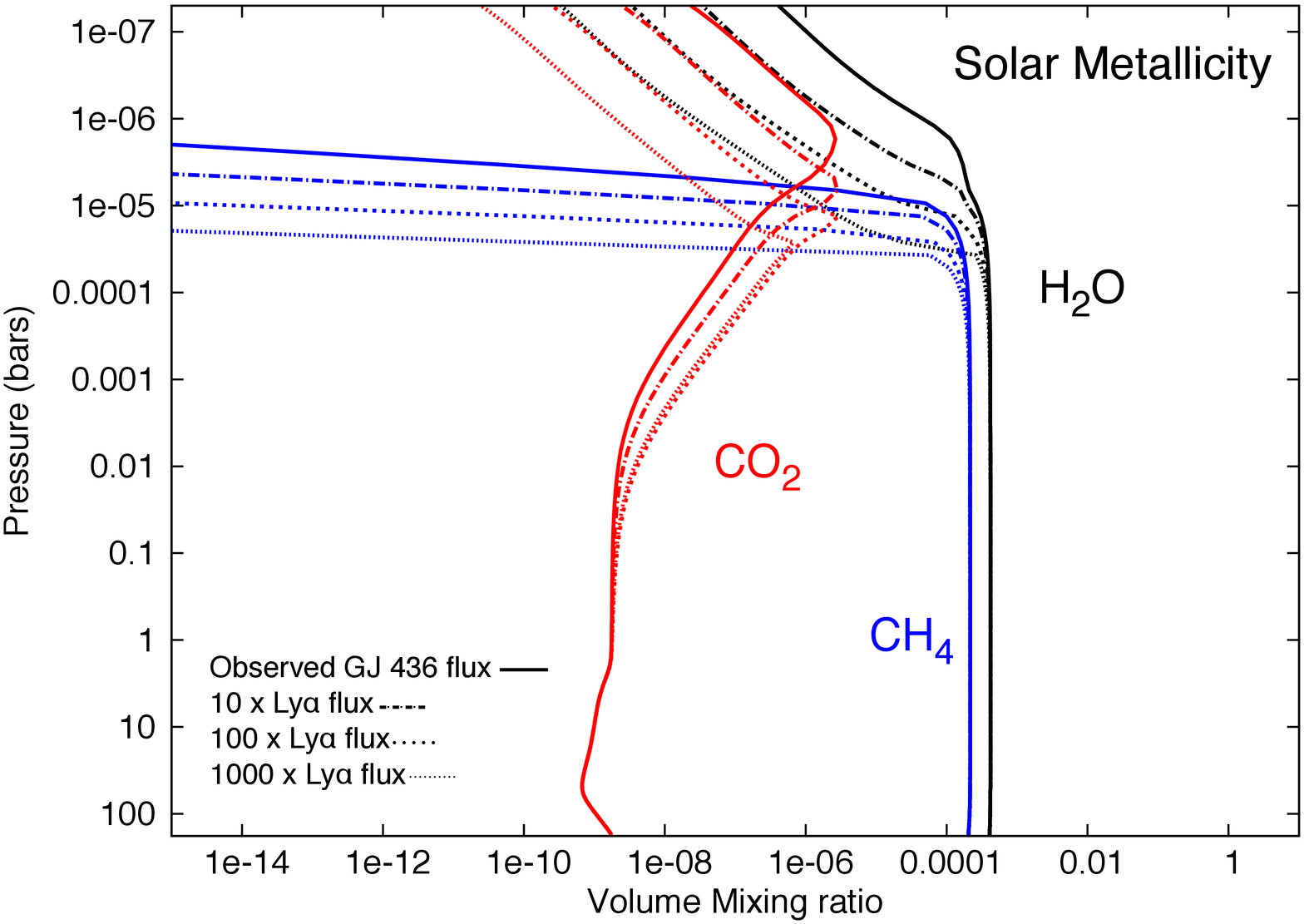}}\subfigure[]{\label{lyman-2}\includegraphics[angle=90,width=.45\textwidth]{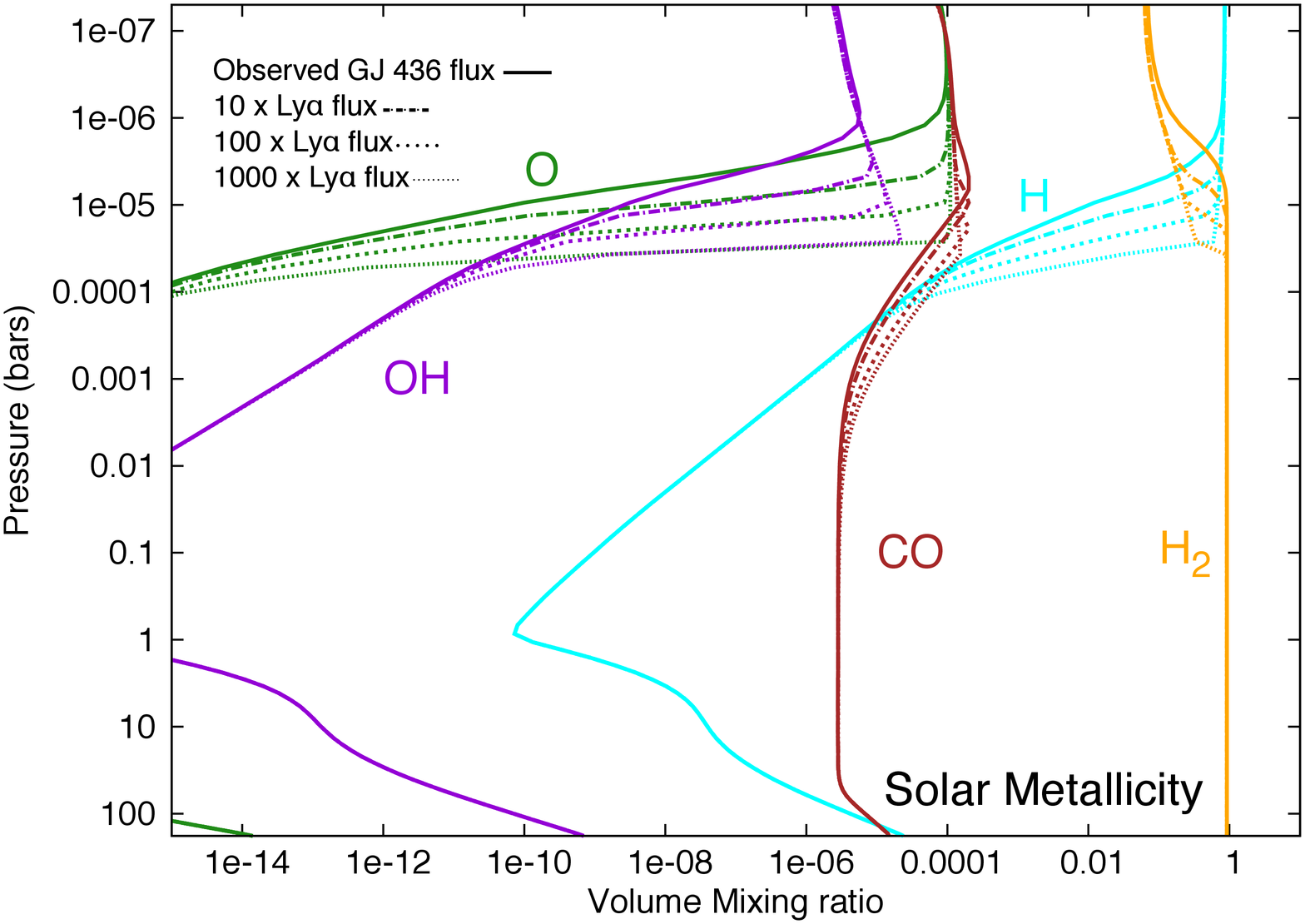}}
\subfigure[]{\label{lyman-metal-1}\includegraphics[angle=90,width=.45\textwidth]{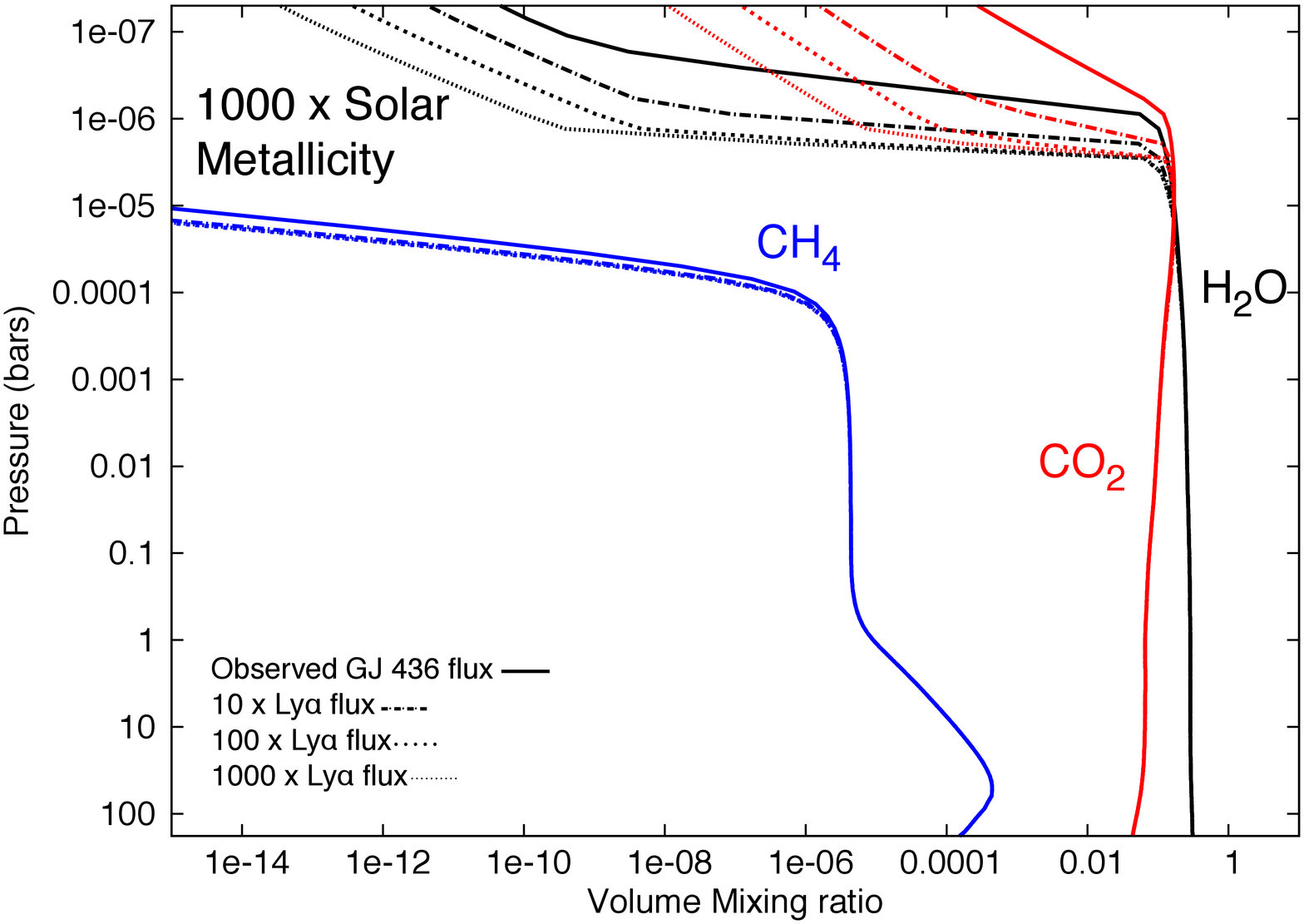}}\subfigure[]{\label{lyman-metal-2}\includegraphics[angle=90,width=.45\textwidth]{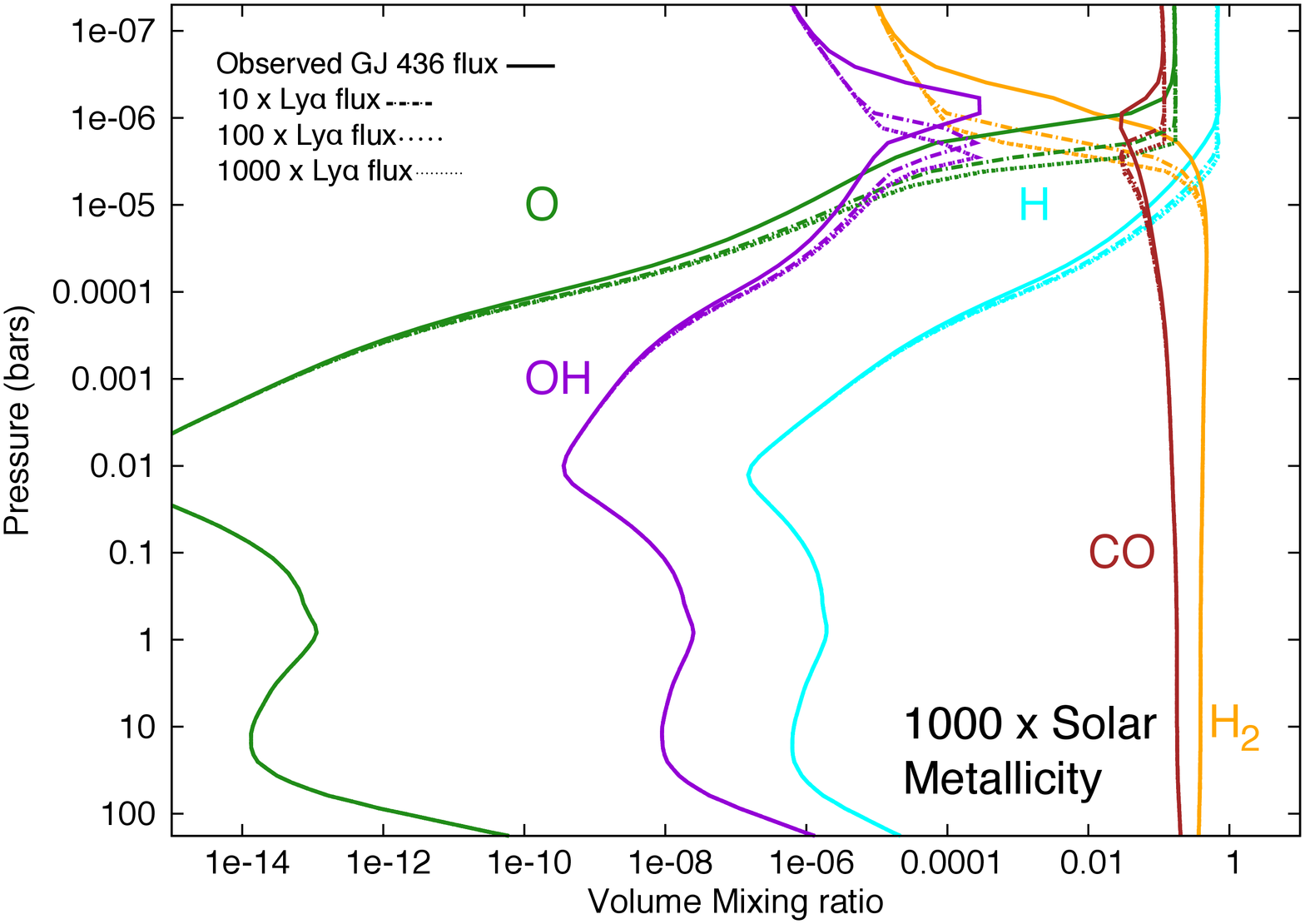}}
 \end{center}
  \caption{Mixing ratios vs. pressures for GJ 436b for solar (top
  panels) and 1000 x solar metallicity (bottom panels) 
  atmospheres. Left panels (Figs. \ref{lyman-1} and \ref{lyman-metal-1}) show
  gases affected by Ly$\alpha$ flux: H$_2$O (black), CH$_4$ (blue),
  CO$_2$ (red) and their concentrations for different Ly$\alpha$ flux
  conditions: 1000 x Ly$\alpha$ flux (small dotted), 100 x Ly$\alpha$
  (dotted), 10 x Ly$\alpha$ (dots and dashes) and 1 x the 
reconstructed Ly$\alpha$ flux (solid line). Right panels
  (Figs. \ref{lyman-2} and \ref{lyman-metal-2}) show the H, H$_2$, O, CO 
and OH mixing ratios.}
\label{photo}
\end{figure*}

In Fig. \ref{photolysis}, we show how the photolysis rates of
  the molecules most susceptible to dissociation by Ly$\alpha$ flux
  (H$_2$O, CO$_2$ and CH$_4$) change in the atmosphere for different
  irradiation scenarios: GJ 436b irradiated by 1000 x Ly$\alpha$,
  100 x Ly$\alpha$, 10 x Ly$\alpha$, and 1 x Ly$\alpha$, where n x
  Ly$\alpha$ is the reconstructed Ly$\alpha$ flux for GJ~436b
  multiplied by the factor n [see
  Fig. \ref{high}]. The photolysis rate of species $i$ (r$_i$) is proportional to the concentration of the species ($n_i$) and the photodissociation coefficient ($J_i$, shown in Fig. \ref{photolysis-frequency}) which depends only on the flux ($F$) and the cross section of the species ($\sigma_i$), as shown in equations (\ref{rates}) and (\ref{coef}) \citep{yung}:

\begin{equation}\label{rates}
r_i(z)=J_i(z)~n_i(z),
\end{equation}
\begin{equation}\label{coef}
J_i(z)=\int{\sigma_i(\lambda)~F(z,\lambda)~d\lambda}.
\end{equation}
The H$_2$O concentration is higher than CH$_4$ and CO$_2$ (see
Fig. \ref{lyman-1}) and, therefore, has higher photolysis
rates. H$_2$O absorbs most of the FUV radiation, becoming optically
thick to radiation when
$\lambda<2000$~\AA$~$at $\sim0.08$~bars. Lower photolysis rates at 
higher pressures are due to radiation at longer wavelengths 
($2000<\lambda<2400$~\AA). 

Fig. \ref{photolysis} shows that high values of the Ly$\alpha$ flux 
(1000, 100, and 10 x Ly$\alpha$) lead to more radiation at higher
pressures in the atmosphere, thereby increasing the H$_2$O, CO$_2$ 
and CH$_4$ photolysis rates.

Figs. \ref{lyman-1} -- \ref{lyman-2} show mixing ratios vs. pressure of
the most abundant species in GJ 436b's atmosphere with solar
composition for four scenarios: 1000 x Ly$\alpha$ (small dotted), 100
x Ly$\alpha$ (dotted), 10 x Ly$\alpha$ (dots and dashes) and 1 x the
reconstructed Ly$\alpha$ flux (solid line). For all  species, mixing
ratios in the four cases are the same for higher pressures, where
photodissociation processes are not efficient, but start to deviate
from equilibrium values when photodissociation of molecules occurs, 
mostly in the upper observable atmosphere (P$<10^{-3}$bars). 

Since GJ 436b is a cool planet with a T$_{\rm eq}\sim 640$~K (assuming
that the albedo=0),
CH$_4$ is the most abundant carbon compound up to $10^{-4}$ bars for
solar composition. At lower pressures, it is oxidized, and its
abundance decreases rapidly. The CH$_4$ photolysis rate changes
significantly with increasing Ly$\alpha$ flux 
(see Fig. \ref{photolysis}), as shown by the difference between the 
two extreme Ly$\alpha$ fluxes  (1000 and 1 x Ly$\alpha$) 
being two orders of magnitude at $5\times10^{-5}$ bars, leading to 
different mixing ratios at lower pressures. 

H$_2$O, which is the most abundant gas after He and H$_2$, starts to 
dissociate at $10^{-4}$~bars for 1000 x Ly$\alpha$, at 
$5\times10^{-5}$~bars for 100 x Ly$\alpha$, at $10^{-5}$~bars for 
10 x Ly$\alpha$ and at $5\times10^{-6}$bars for the reconstructed 
Ly$\alpha$ flux. 

The photolysis of H$_2$O affects the mixing ratios of other
species in the atmosphere such as O, OH and H which increase with increasing
H$_2$O photolysis. H$_2$O dissociation is followed by the
destruction of H$_2$ because of a reaction with OH. As a consequence
of these reactions, a large amount of H is created and H$_2$ is
destroyed with increasing Ly$\alpha$ flux. H replaces
H$_2$ as the most abundant gas in the atmosphere at pressures
P$<5\times10^{-5}$~bars in all cases. OH increases with the Ly$\alpha$
flux because of the H$_2$O photolysis at $\sim 10^{-5}$~bars. Note that CO photolysis is not considered because its photolysis is driven by EUV radiation which is not included in the model. Atomic O is produced from H$_2$O photolysis. At $10^{-5}~$bars, its mixing ratio is $10^{-4}$ for 1000 x Ly$\alpha$ and $10^{-10}$ for 1 x Ly$\alpha$ flux.

CO and CO$_2$ show different behaviors compared to their chemistry in
hot Jupiters' atmospheres, because CO is the dominant carbon compound
in hot exoplanet atmospheres, whereas CH$_4$ is dominant in cooler
planets. Self-shielding by CH$_4$ and other effects lead to differences in the chemistry and pressure
where self-shielding occurs \citep{li11}. The
CO-mixing ratio increases because of photochemistry at $\sim0.01$~bars, increasing up to 2 orders of magnitude at $\sim10^{-5}$~bars. CO$_2$ is formed after H$_2$O photolysis, with a local maximum at $\sim10^{-4}$~bars for 1000 x Ly$\alpha$, at $5\times10^{-5}$~bars for 100 x Ly$\alpha$, at $8\times10^{-6}$~bars for 10 x Ly$\alpha$ and at $10^{-6}$~bars for 1 x Ly$\alpha$. 

Our results for the dominant carbon and oxygen species in GJ 436b for
1 x Ly$\alpha$ flux are in good agreement with previous results by
\citet{li11} and \citet{mo13}, with small differences due to
differences in the UV fluxes used, cross sections adopted, 
chemical schemes and thermal profiles. 

\begin{flushleft}
\subsubsection{Atmospheres with high metallicities}\label{1000xsolar}
\end{flushleft}

High metallicities are expected in some mini-Neptune atmospheres
because of the enrichment inferred from interior modeling
\citep{ba08,fi09,ne10} and from population synthesis models
\citep{fo13}. As a test case, we compute simulations for the
atmosphere of GJ 436b assuming 1000 x solar metallicity.

An atmosphere with 1000 x solar metallicity has lower hydrogen and
helium and increased carbon and oxygen abundances
\citep{mo13}. Fig. \ref{photolysis-metal} shows the photolysis rates
of CH$_4$ (blue), H$_2$O (black) and CO$_2$ (red) with the
reconstructed Ly$\alpha$ flux of GJ 436 for a solar metallicity
atmosphere (solid lines) and an atmosphere with 1000 x solar
metallicity (dotted lines). H$_2$O shows the highest photolysis rates
for both compositions. The CO$_2$ photolysis rate is smaller than
CH$_4$ for solar composition, 
but it increases for the 1000 x solar model, becoming larger than H$_2$O at $P<1\times10^{-6}$~bars. CH$_4$ has very low photolysis rates, especially for the case of an enriched atmosphere because it is shielded by H$_2$O and CO$_2$.

Figs. \ref{lyman-metal-1} -- \ref{lyman-metal-2} show mixing ratios
vs. pressure for H$_2$O, CO$_2$, CH$_4$ and for H, H$_2$, O, CO and
OH, respectively, for an atmosphere of 1000 x solar metallicity and
four scenarios: 
1000 x Ly$\alpha$ (small dotted), 100 x Ly$\alpha$ (dotted), 
10 x Ly$\alpha$ (dots and dashes) and 1 x Ly$\alpha$ flux for GJ 436
(solid line). The photolysis rates (Fig. \ref{photolysis-metal}) are
reflected in the mixing-ratio profiles of H$_2$O, CO$_2$, CH$_4$ and
those of the other molecules affected by H$_2$O dissociation. Since
H$_2$O and CO$_2$ compete for photons, 
increased dissociation of CO$_2$ coincides with decreased dissociation 
of H$_2$O compared to a solar metallicity atmosphere. As a
consequence, H, H$_2$, O, CO and OH show less change for different
irradiation cases. Dissociation of H$_2$O and CO$_2$ starts at
$5\times10^{-6}$~bars in all cases, 
but  mixing ratios are smaller for larger Ly$\alpha$ fluxs. CH$_4$ 
shows very little change in the four scenarios because it is shielded 
by H$_2$O and CO$_2$.

\subsection{Absorption of Ly$\alpha$ radiation in the extended H atmosphere}\label{discussion}

Some hot extrasolar planets have an extended atmosphere
  consisting primarily
  of atomic H which could efficiently absorb Ly$\alpha$
  radiation, as shown by observations \citep{vm03,lec10,eh12} and
  theoretical models \citep{ko10,lav11}. Some absorption of
  Ly$\alpha$ photons has even been detected in GJ 436b's atmosphere
  \citep{ku14} which increases the importance of studying the
  absorption of Ly$\alpha$ radiation in an exoplanet's atmosphere and
  the effects on 
  photochemistry. The exoplanet thermosphere is characterized by an
  increase in temperature \citep{hu12} which may affect the
  resulting atmospheric composition in the upper atmosphere (P
  $<\approx10^{-7}$~bars) for some planets with extremely high
  temperatures. For those cases, thermochemical processes dominate the
  chemistry in the upper atmosphere, and the Ly$\alpha$ flux plays no role
  in defining the atmospheric composition \citep{lav11}. On the other
  hand, there are cooler planets, such as GJ~436b, for which the
  incident Ly$\alpha$ flux may be partially absorbed in the atmosphere,
  but the effects of this decreased
  flux on the resulting composition remain poorly
  understood. \citet{lav11} studied this problem and performed
  photochemical calculations taking into account the exoplanets'
  thermosphere for three planets, including GJ~436b, where they found
  that Ly$\alpha$ flux plays a small role in the photochemistry. In the
  present work, we use the FUV flux taken from recent observations
  \citep{fr13} which is at least one order of magnitude larger than
  the FUV flux used in \citet{lav11}. We assume that this Ly$\alpha$ 
  radiation is not completely
  absorbed and therefore plays an important role in the photochemistry
  described in Section \ref{results}. It may be possible, nevertheless,
  that some Ly$\alpha$ radiation is absorbed. In this section,
  we explore the photochemistry effects of the absorption of Ly$\alpha$ flux
  in the atmosphere of GJ~436b. Since a self-consistent
  model of thermal structure, photochemistry and thermosphere is beyond
  the scope of this paper, we use a simplified model in which a
  portion of the stellar Ly$\alpha$ radiation is absorbed in the extended H
  atmosphere. We consider the three possible absorption scenarios shown
  in Fig. \ref{low}: 0.1 x Ly$\alpha$,  0.01 x Ly$\alpha$ and 0.001 x
  the reconstructed Ly$\alpha$ flux.

\begin{figure*}
  \begin{center}
\subfigure[]{\label{lyman-1-abs}\includegraphics[angle=90,width=.45\textwidth]{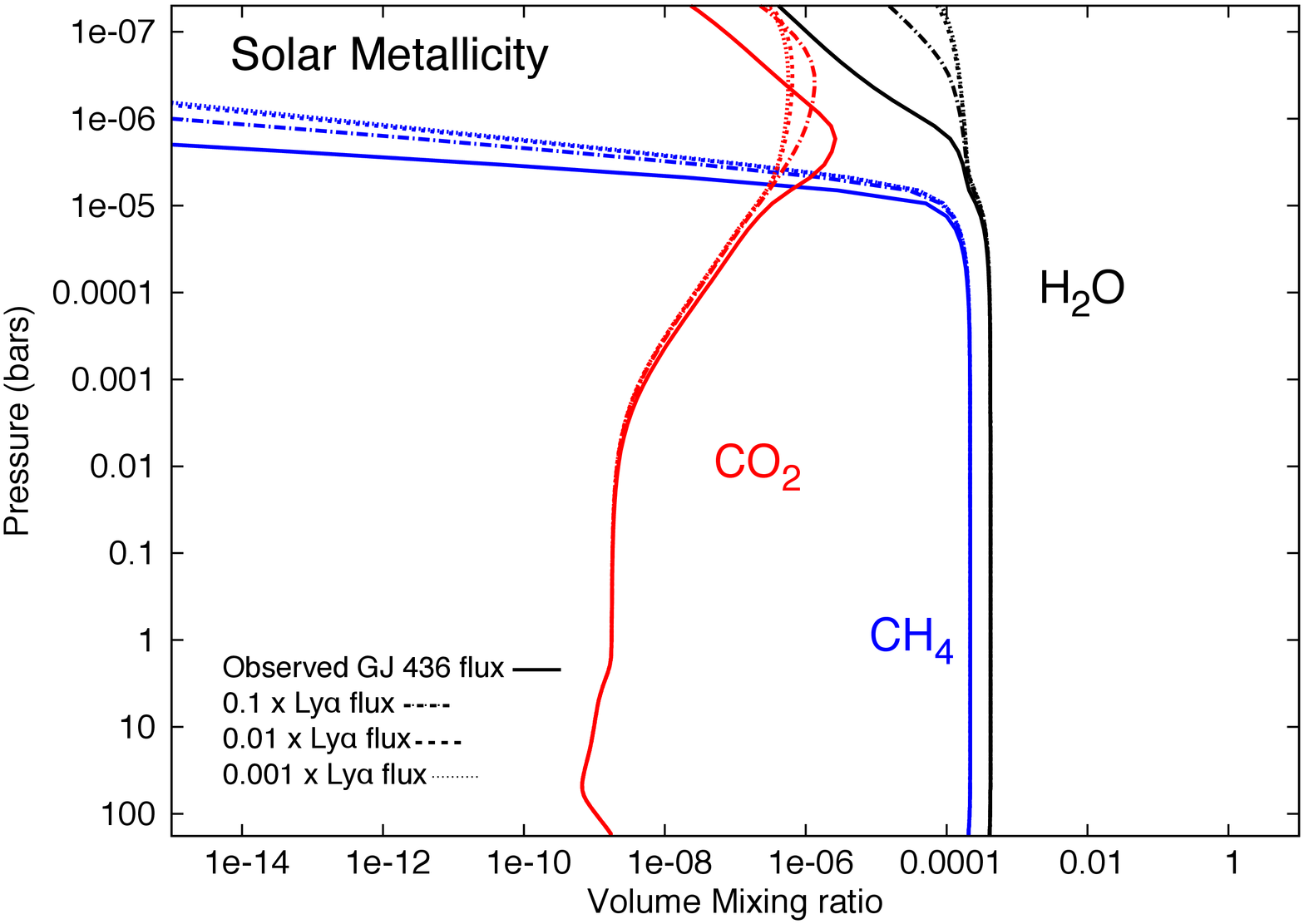}}\subfigure[]{\label{lyman-2-abs}\includegraphics[angle=90,width=.45\textwidth]{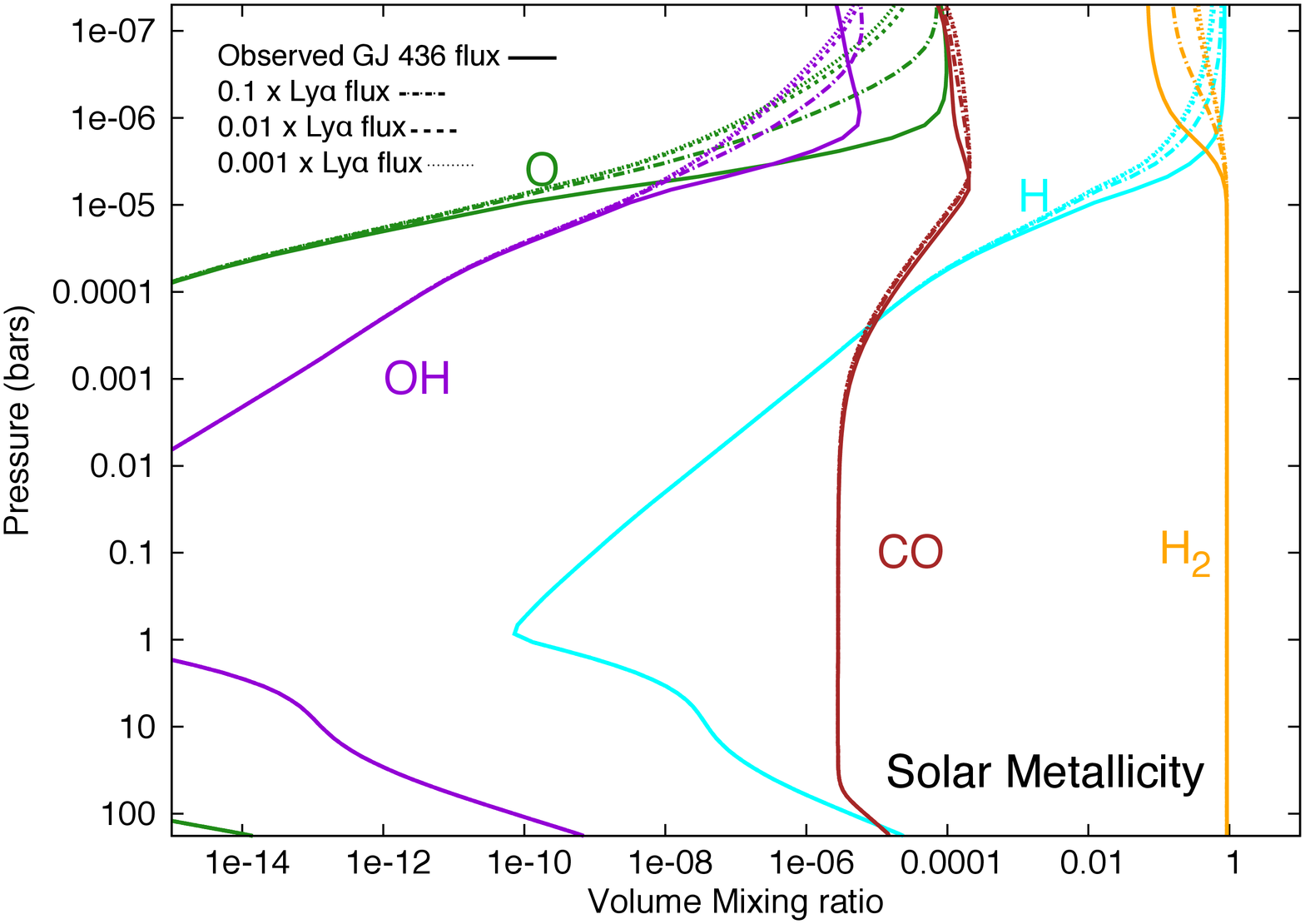}}
\subfigure[]{\label{lyman-metal-1-abs}\includegraphics[angle=90,width=.45\textwidth]{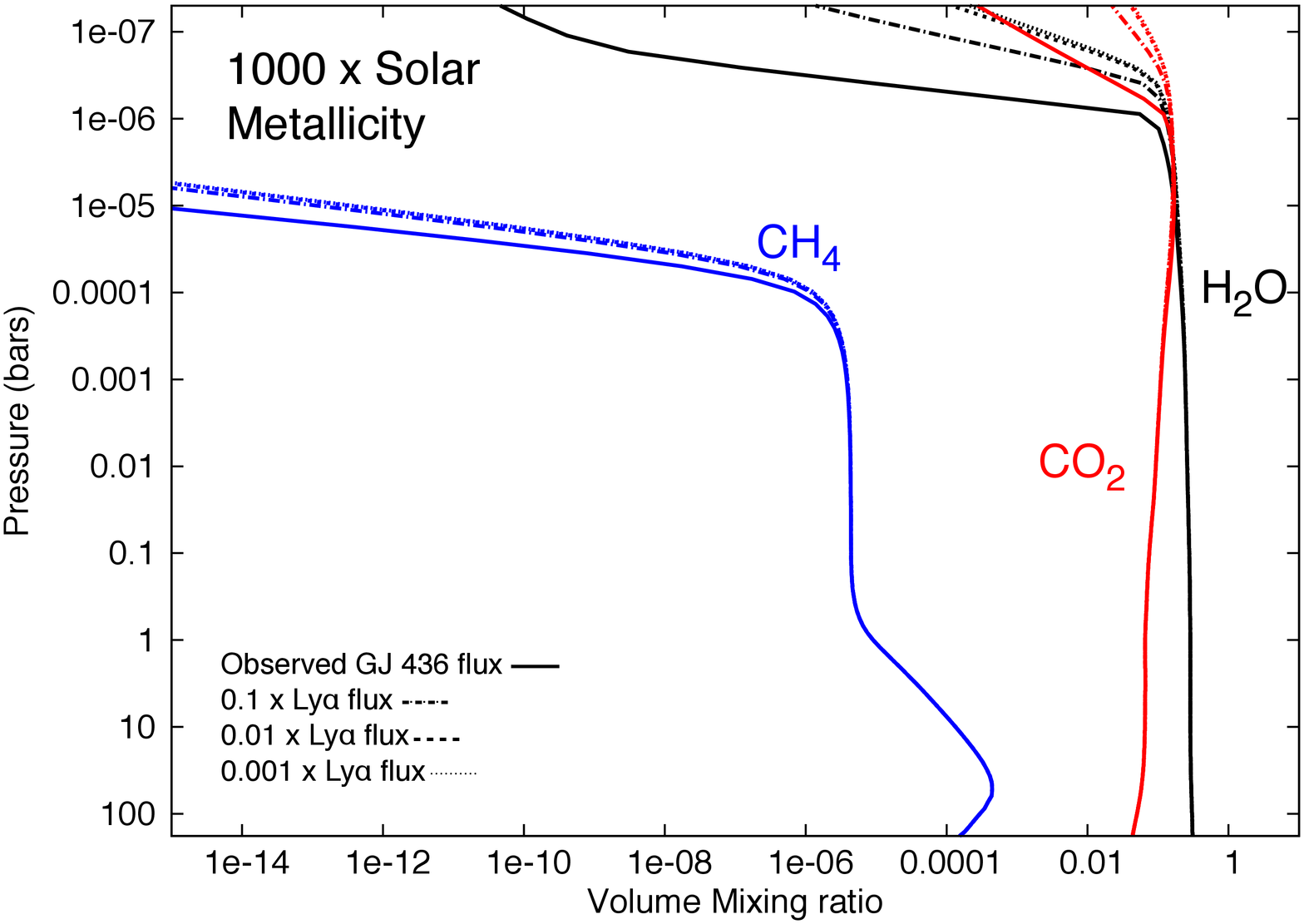}}\subfigure[]{\label{lyman-metal-2-abs}\includegraphics[angle=90,width=.45\textwidth]{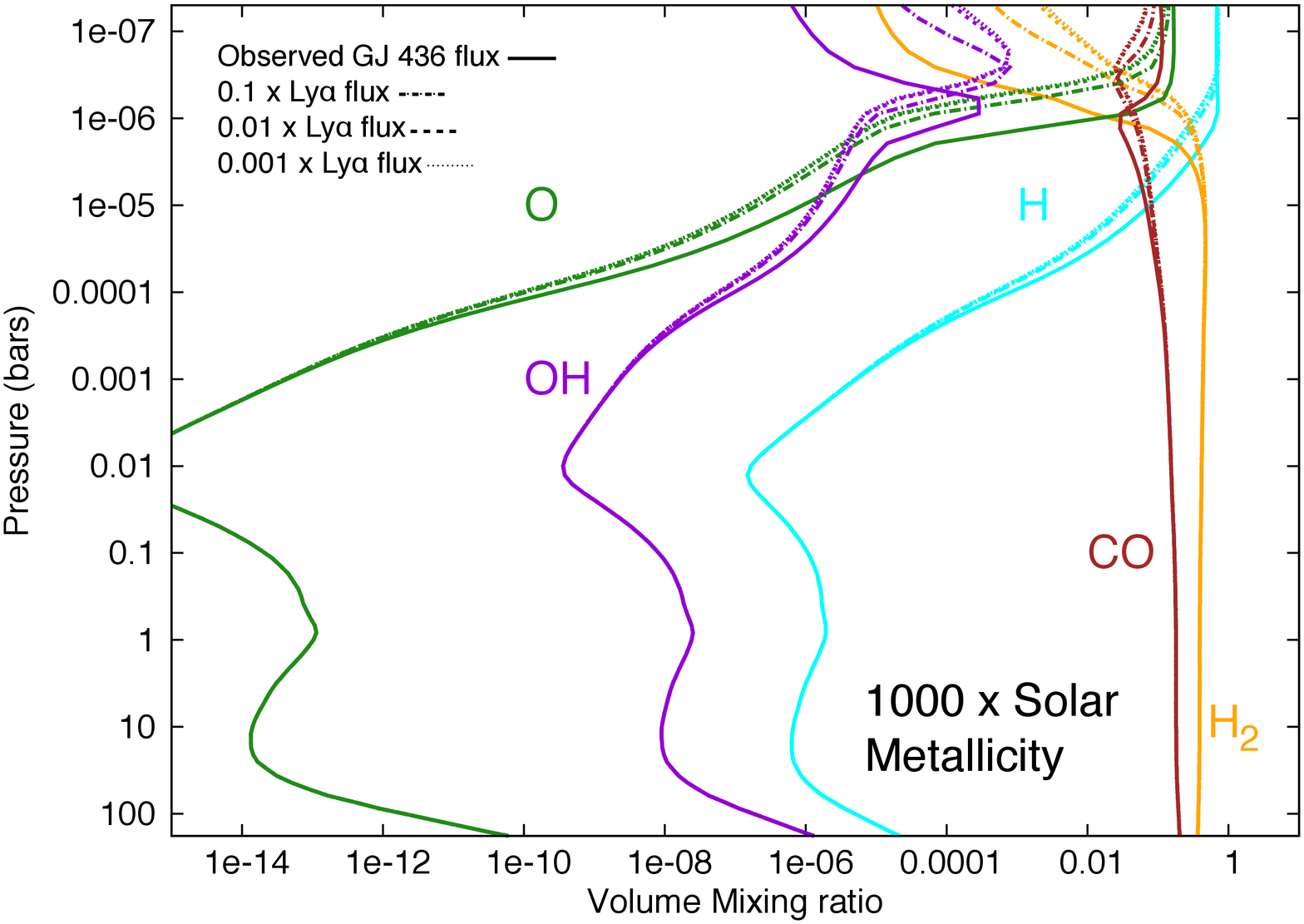}}
 \end{center}
  \caption{Photochemical mixing ratios vs. pressure for an atmosphere with solar composition (upper panels) and 1000 x solar metallicity (lower panels) for different Ly$\alpha$ absorption scenarios: 1 x Ly$\alpha$ (solid), 0.1 x Ly$\alpha$ (dots and dashes), 0.01 x Ly$\alpha$ (dashed) and 0.001 x Ly$\alpha$ flux (dotted line).}
\label{photo-Lyabs}
\end{figure*}

In Fig. \ref{photo-Lyabs} we show the mixing ratios
  of H$_2$O, CH$_4$, CO$_2$, O, OH, H, CO and H$_2$ as a function of
  the pressure in GJ 436b's atmosphere. We investigate the results of
  adopting solar composition (Fig. \ref{lyman-1-abs} --
  \ref{lyman-2-abs}) as well as 1000 x solar metallicity in the
  atmosphere (Fig. \ref{lyman-metal-1-abs} -- \ref{lyman-metal-2-abs})
  in the three explored scenarios 
(0.1 x Ly$\alpha$, 0.01 x Ly$\alpha$ and 0.001 x Ly$\alpha$). Since 
the atmosphere is exposed to reduced Ly$\alpha$ radiation, the 
photolysis rates are reduced in all cases compared to a deeper
  penetration of Ly$\alpha$ stellar flux (as adopted in Section
  \ref{results}). Decreased dissociation of H$_2$O leads to an
  increase in its mixing ratio which is the highest in the case of
  extreme absorption (0.001 x Ly$\alpha$). For the case of solar
  composition, the mixing ratio of H$_2$O at $5\times10^{-6}$~bars is
  $\sim2\times10^{-6}$ for 1 x Ly$\alpha$ and $\sim2\times10^{-4}$ for
  0.001 x Ly$\alpha$ (Fig. \ref{lyman-1-abs}), and for 1000 x solar
  composition, it is $\sim1\times10^{-7}$ for 1 x Ly$\alpha$ and
  $0.05$ for 0.001 x Ly$\alpha$ at the same pressure
  (Fig. \ref{lyman-metal-1-abs}). These changes in the H$_2$O-mixing
  ratios lead to changes in O, OH, H, CO and H$_2$ 
(see Fig. \ref{lyman-2-abs} -- \ref{lyman-metal-2-abs}).
  
For a solar composition atmosphere, the increase in the CO$_2$-mixing
ratio at $\sim1\times10^{-5}$~bars is a consequence of H$_2$O
photolysis, but  the CO$_2$-mixing ratio decreases when there is less 
H$_2$O dissociation for higher absorption of Ly$\alpha$ flux.  The
change between the extreme cases (1 x Ly$\alpha$ and 0.001 x
Ly$\alpha$ flux) is one order of magnitude at $5\times10^{-6}$
bars. For solar composition and 1000 x solar metallicity atmospheres,
the CO$_2$-mixing ratio at lower pressures (P$<5\times10^{-6}$~bars)
is dominated by its own dissociation, therefore the CO$_2$-mixing
ratio increases for higher Ly$\alpha$ flux absorption. The difference
in the CO$_2$-mixing ratios for the extreme cases (1 x Ly$\alpha$ and
0.001 x Ly$\alpha$) at $\sim1\times10^{-7}$ bars is one order of
magnitude for a solar metallicity atmosphere and two orders of
magnitude for a 1000 x solar metallicity atmosphere.

CH$_4$ dissociation is caused mainly by radiation around
1300~\AA$~$(see Fig. \ref{cross-section}). Since this molecule is also
shielded by H$_2$O (solar composition) as well as by H$_2$O and CO$_2$
(1000 x solar metallicity), its mixing ratio does not change
significantly with increasing  absorption of Ly$\alpha$ flux. 
 
The absorption of Ly$\alpha$ radiation is important for the
photochemistry in these exoplanet atmospheres. Different absorption
scenarios lead to different mixing ratios and, therefore, it is
necessary to know the amount of flux absorbed to know the
effect on the photochemistry. Nevertheless, we notice that very strong
absorption (0.001 x Ly$\alpha$ flux) has only a small effect on the photochemistry (compared to the case of 0.01 x Ly$\alpha$ flux) because the dissociation of molecules is mainly due to radiation at other wavelengths.

\section{CONCLUSION}\label{conclusion}

Ly$\alpha$ radiation changes mini-Neptunes' upper atmospheric
chemistry significantly. We explore the effect of Ly$\alpha$-driven
photochemistry for atmospheres with
different metallicities, comparing solar composition and a 1000 x solar
composition. Focusing on GJ 436b as an example, we calculate the
thermal structure and chemistry including equilibrium and
disequilibrium chemistry (molecular diffusion, vertical mixing, and
photochemistry). We use direct observations of the UV and the
  reconstructed Ly$\alpha$ flux for the host star GJ 436
  \citep{fr13}. We explore the effects on the planet's atmosphere 
of increasing the incident 
Ly$\alpha$ flux by factors of 10, 100 and 1000 as well as the case
where the Ly$\alpha$ flux is absorbed in the extended H atmosphere by factors of 0.1, 0.01 and 0.001.

For solar composition atmospheres, our results show that the mixing
ration of H$_2$O is most affected by Ly$\alpha$ radiation as the
H$_2$O photolysis rate strongly depends on the Ly$\alpha$ flux even at
pressures as large as $\sim 0.08$~bars. The H$_2$O-mixing ratios
change by up to five orders of magnitude between the cases of 1000 x
Ly$\alpha$ and 0.001 x Ly$\alpha$. The reconstructed Ly$\alpha$ flux thereby
significantly changes the upper atmospheric chemistry and the
resulting observable spectral features. H$_2$O is one of the most
abundant gases in the atmosphere, absorbing much of the Ly$\alpha$
flux and shielding other species like CH$_4$. Changes in the H$_2$O
photolysis rates also affect other molecules whose mixing ratios are
largely affected by H$_2$O dissociation: for example, CO$_2$ changes
by up to 3 orders of magnitude, OH by up to 5 orders of magnitude, H
by up to 4 orders of magnitude, O by up to 6 orders of magnitude, 
H$_2$ by 1 order of magnitude, and CO by less than 1 order of
magnitude between the extreme cases of 1000 x Ly$\alpha$ and 
0.001 x Ly$\alpha$ flux.

Because of the high abundance of CO$_2$ in high metallicity
atmospheres, CO$_2$ competes with H$_2$O for energetic FUV
photons. The CO$_2$ photolysis rate is largely affected by the
Ly$\alpha$ flux, and therefore its mixing ratio changes by up 
to 4 orders of magnitude for the extreme irradiation scenarios we have
explored. These two molecules absorb most of the Ly$\alpha$ radiation,
thereby shielding CH$_4$. The smaller effect on H$_2$O also leads to
smaller changes in the abundance of those other molecules for which  
mixing ratios in the upper atmosphere strongly depend on H$_2$O photolysis. 

Our models show that Ly$\alpha$ radiation from the host star affects
mini-Neptune atmospheres significantly at low pressures and cannot be
ignored in atmospheric modeling. Ly$\alpha$ radiation affects the
photochemistry of important gases in the upper atmosphere and,
therefore, also the resulting observable spectral features of
mini-Neptunes. Observations of the UV flux from a wide range of stars
as well as studies of the absorption of this radiation in the
exoplanet thermospheres are essential for realistic interpretations 
of planetary spectra.

\begin{flushleft}
\section*{ACKNOWLEDGEMENTS}
\end{flushleft}

We would like to thank Ravi Kopparapu, James Kasting, Dimitar
Sasselov, Kevin France and Yan Betremieux for useful
discussions. Special thanks to Julianne Moses, for fruitful
discussions and providing the thermal structure of high-metallicity GJ
436b atmosphere. YM and LK acknowledge DFG funding ENP Ka 3142/1-1 and
the Simons Foundation. JLL acknowledges support from the Space
Telescope Science Institute.
This work has made use of the MUSCLES M dwarf UV radiation field database.

\end{document}